\documentclass[prl,preprint,12pt]{revtex4}
\usepackage{epsfig,wrapfig}
\begin{document}
\title{Structural Signatures of Vitrification in  Hard Core Fluids}
\author{Sanat K. Kumar\footnote{E-mail:kumar@rpi.edu}, Shekhar Garde}
\affiliation{Department of Chemical Engineering, Rensselaer Polytechnic Institute, Troy, NY}
\author{Jack F. Douglas, Francis W. Starr}
\affiliation{Polymers Division and Center for Theoretical and Computational Materials Science, 
National Institutes of Standards and Technology, Gaithersburg, MD}

\begin{abstract}
Computer simulations of hard spheres and disks
are used to estimate the most probable
cavity size, $\xi_{\rm cavity}$, and a ``rattle'' size, $\xi_{\rm rattle}$, over which
a particle can translate holding all other particles fixed. Both of these measures
of free volume appear to extrapolate to zero at the random close 
packed density, $\rho_{\rm rcp}$, close to the density
where extrapolations of the viscosity diverge. %Over a limited density
%range relevant to experiments these
%free volume measures vary nearly linearly with $\rho$ and 
%extrapolate to zero at the experimentally determined glass transition.
We also identify the onset of caging as the density at which 
cavities cluster.
%These results  
%emphasize the importance of microscopic liquid structure, especially
%free volume, to vitrification. 
These results suggest that vitrification in hard core fluids can be viewed as a geometrical phenomenon, and that
{\em local} free volume measures can identify the location of 
the onset of liquid-like dynamics, the complex dynamics of caged liquids,
and vitrification.
\end{abstract}

\maketitle
Ensembles of hard spheres have long been utilized as prototypical models for liquids
and solids. 
%However, many questions remain about the conditions under which these material
%states emerge, and the structural characteristics that define them. 
Since these systems
can be defined purely by geometry, it seems reasonable to understand them in terms
of void or ``free" volume. Thus, a class of models, termed ``free volume" models have
been very popular in describing the transition of hard core fluids from the gas to the
liquid state, from the liquid to the crystal and from the liquid to the glassy state
\cite{doolittle,cohen,cohengrest,starr,giaquinta}. Inspite of the popularity of these
ideas, the molecular connection between ``free
volume" and dynamic transitions in these fluids remains unclear \cite{finney}.
%Here we seek to understand this connection for the prototypical case of hard core liquids.
Previous molecular dynamics simulations \cite{hoover,henderson,alder} have identified many dynamic signatures of the
vitrification of hard core fluids and it is our goal to enunciate any underlying 
structural indicators. 
%Thus,
%we utilize Monte Carlo computer simulations to quantify the density dependence
%of well-defined measures of the cavity and rattle volume in the prototypical
%cases of hard sphere and hard disk fluids. 

We simulate hard spheres and hard disks using the Monte Carlo method and periodic
boundary conditions. The lateral size of the system, $L$, and the number of particles, $N$, are held fixed
in each simulation. We used $L$=100 in two dimensions ($D$=2) 
and up to $L$=60 for $D$=3.
Density, or equivalently volume fraction, is defined as $\rho\equiv\frac{\pi \sigma^2}{4} \frac{N}{L^2}$ in $D$=2
and $\rho\equiv\frac{\pi \sigma^3}{6} \frac{N}{L^3}$ in $D$=3 where
$\sigma$ is the particle diameter. (We report lengths in units
of $\sigma$.) We generate random initial configurations, and thus some
particle overlap was initally unavoidable at the highest densities.
Equilibration was monitored by comparing various system properties, including the pressure and the
chemical potential, to literature estimates (for example
\cite{torquato}). We also confirmed that system did not ``age" as a function
of time over the course of the simulations (10$^6$ attempted moves/particle at all densities).
%We estimate the chemical potential, $\mu$: $\frac{\mu-\mu_{ideal}}{k_BT} = -\ln p_{ins}$.
%Here $\mu_{ideal}$ is the chemical potential of an ideal gas at the same
%density, and $p_{ins}$ is the probability of inserting a molecule into the fluid without
%overlapping any other particle. 

\begin{wrapfigure}{c}{8in}
\mbox{\epsfig{file=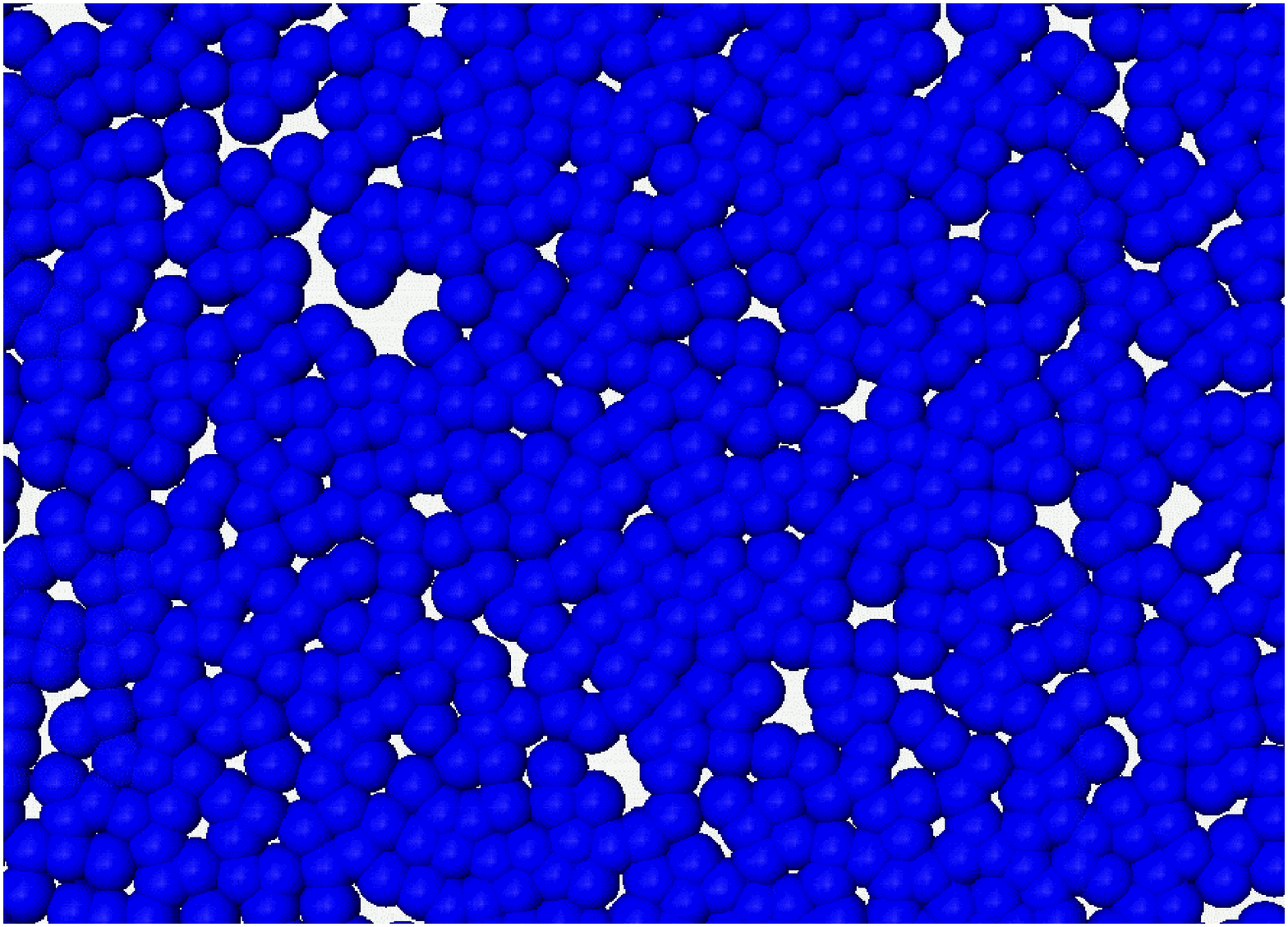, width=3.in}}
\mbox{\epsfig{file=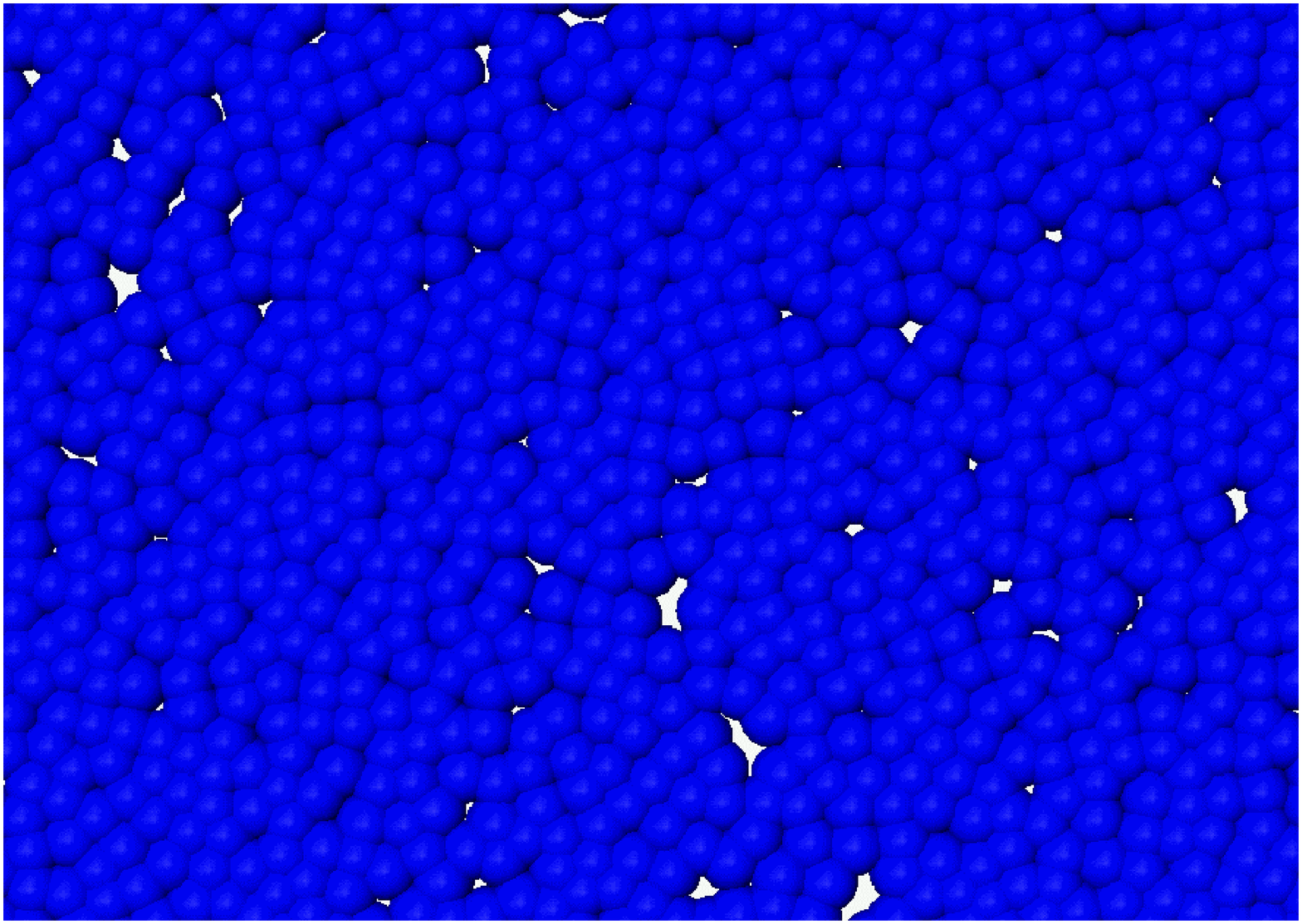, width=3.in}}

\vspace{40mm}
\mbox{\epsfig{file=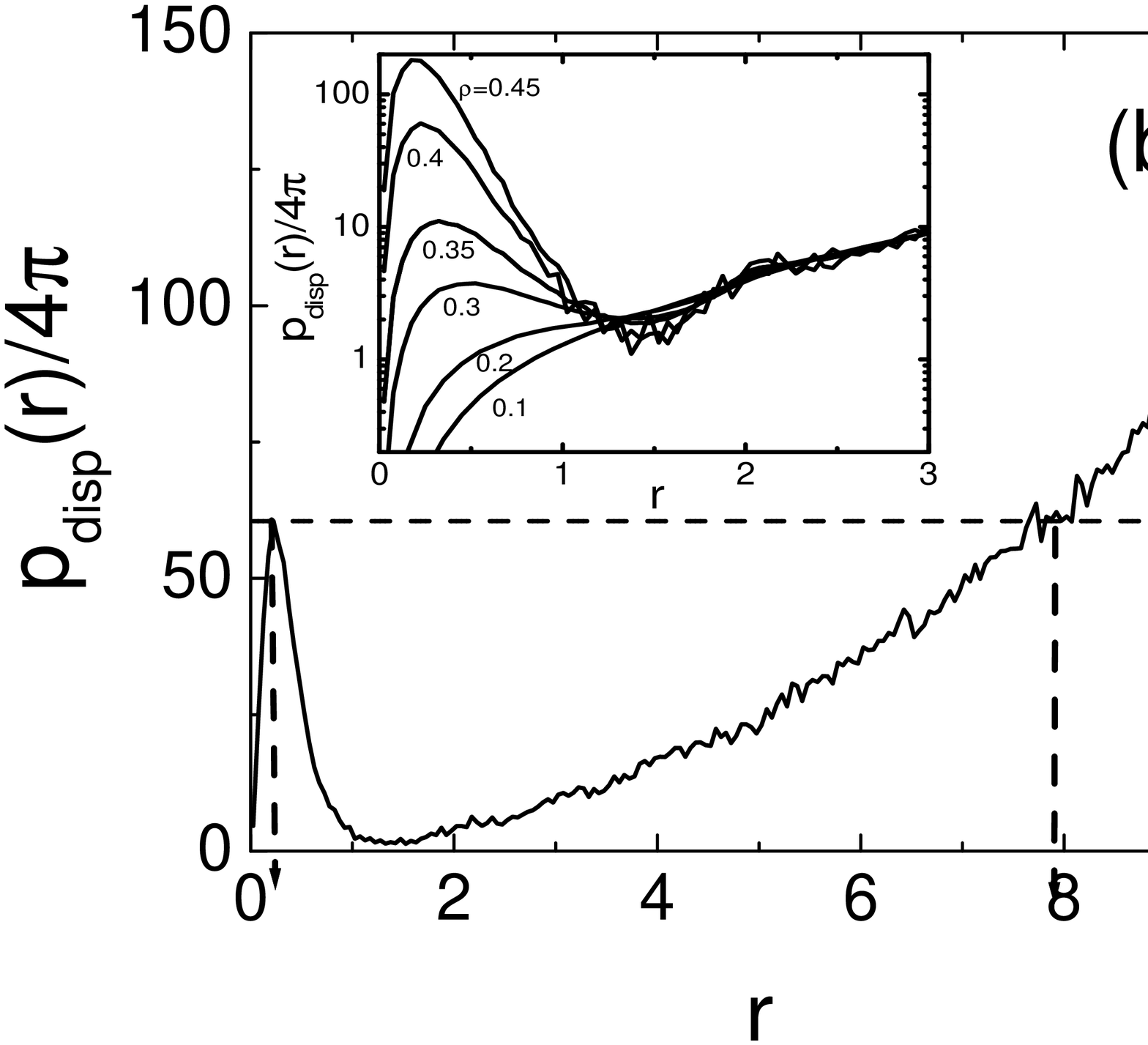, width=6in}}
\vspace{-40 mm}
\caption{}
\label{fig1}
\label{snapshot}
%\end{center}
\end{wrapfigure}

Representative particle configurations for relatively low ($\rho$=0.40) and
high density ($\rho$=0.53) hard disc ($D$=2) fluids are shown in Fig. \ref{snapshot}.
Note that although the disks are themselves of diameter 1, we have drawn regions of
exclusion of diameter 2 in this figure. Thus, the unshaded areas correspond to
regions where centers of new particles can be accomodated in the fluid without overlap.
We observe that the ``voids" in this figure become smaller and sparser as the
$\rho$ increases, and it is this effect we quantify in this paper.

We evaluate two characteristic structural size scales using algorithms based
on particle insertions \cite{pratt,reiss}. 
%There have been many previous studies focused on the definition 
%of free volume in hard core fluids \cite{hoover,finney,reiss,pratt,speedyreiss,reiss}.
Following Ref. \cite{pratt}, ``cavity" size distributions are defined by
randomly selecting points in space
and determining the size of the largest spherical cavity that can be inserted about them.
This definition of a ``cavity" is notably different from that
employed in refs. \cite{sastryholes,speedyreiss}.
The distribution function for these maximal spheres displays
a single peak, which is the most probable cavity size, $\xi_{\rm cavity}$.
We expect $\xi_{\rm cavity}$ to be related to
characteristic cavity sizes measured by positron annihilation lifetime spectroscopy,
but the precise relationship has not been conclusively demonstrated.
We find that $\xi_{\rm cavity}$ varies linearly
with $1/\beta\mu$.
Since both $\beta\mu$ and the viscosity of these
fluids are expected to diverge near the random closed pack limit \cite{torquato,russel,woodcock,weitz,chaikin}, a phenomenological
connection between system dynamics and the presence of cavities is implied.
{\em We note that, although the most probable cavity size is expected to go to zero
in this limit, cavities are still expected to exist beyond this point.}

Another useful measure of free volume is the ``rattle volume" which corresponds to the volume explored by a
particle given the constraints of surrounding particles \cite{starr,hoover,henderson}. 
%We evaluate this quantity
%somewhat differently than in past work. 
%\xi_{\rm rattle}$ defines the lateral size of this rattle volume.
Rattle size is evaluated in two equivalent ways.
We consider the equilibrium distribution of acceptable displacements, $p_{\rm disp}(r)$, made
by randomly selected particles while holding all other particles
fixed (see Fig. \ref{fig1}b). The maximum test displacement size is the box size so that all 
possible displacements are sampled. 
Alternatively, we can obtain the same results by considering successful insertions of pairs of spheres of size
unity into the fluid without overlap, and constructing a probability distribution
function for their separations. 
Notably, $p_{\rm disp}(r)$ has a peak at small $r$ values
for $\rho \gtrsim$ 0.25 in $D$=3, and for $\rho \gtrsim$ 0.35 in $D$=2 (see Fig. \ref{fig1}b inset). Since
the chosen particle is hindered by surrounding particles for distances beyond
the peak, we identify the peak as the most probable rattle size, $\xi_{\rm rattle}$.
Past work \cite{starr} suggests that this rattle size is closely related to the
Debye-Waller factor derived from dynamic scattering experiments.

The fact that there is no peak in $p_{\rm disp}(r)$ for $\rho\le$ 0.25 ($D$=3) and $\rho\le$ 0.35 ($D$=2), 
implies that there is no well-defined rattle volume in this density range, i.e.,
particles can freely move over the whole volume, as expected for a gas. For higher
densities, however, our results suggest that rattle sizes are no longer ``extensive" (i.e.,
scales with the system size),
suggesting that holes are localized [see Fig. \ref{snapshot}]. This is 
consistent with the fact that $\rho \approx$ 0.26
has previously been assigned as the onset of the ``fluid" region for hard sphere dynamics ($D$=3), 
where the free volume no longer percolates \cite{giaquinta}. %it is reasonable
%that $\xi_{\rm rattle}$ is no longer defined at lower densities.
%For all smaller densities $\xi_{\rm rattle}$ must scale with system size, i.e., there
%would be no peak at small $r$ values. 
Hence, $\xi_{\rm rattle}$ identifies the region where hard fluid transport becomes
liquid-like. Note that this density is
much less than the density at the onset of ``caging" in the dynamics, i.e.,
$\rho$ = 0.53 for $D$=3, as we shall discuss below \cite{cagingheuer}. 
To further test the utility of $\xi_{\rm rattle}$, we consider the
Lindemann criterion as applied to an amorphous phase \cite{wolynes}. 
This criterion requires that the ratio
of the mean rattle distance to the interparticle distance be $\approx$ 0.1-0.15 at the
transition density. Specifically, for both spheres and disks we find that
$\xi_{\rm rattle} \, \rho^{1/D}\approx$ 0.08$\pm$0.02 at the crystallization density, 
consistent with this generalized Lindemann criterion. 
These two results support the utility of these free volume ideas.

\begin{wrapfigure}{c}{8in}
\vspace{30 mm}
\mbox{\epsfig{file=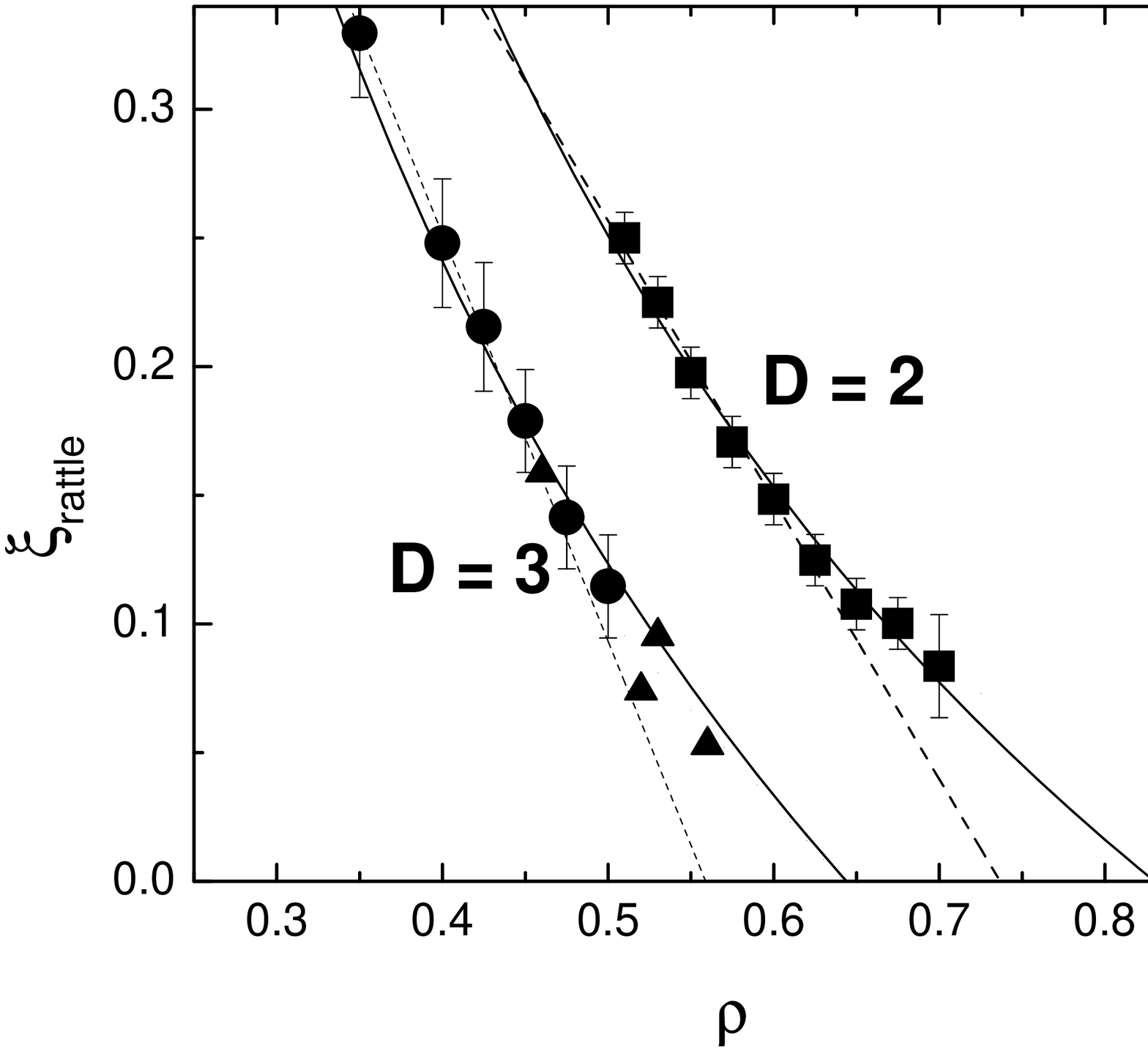, width=8in}}
\vspace{-40 mm}
\caption{}
\label{fig3}
%\end{center}
\end{wrapfigure}

\begin{wrapfigure}{c}{7in}
\vspace{30 mm}
\mbox{\epsfig{file=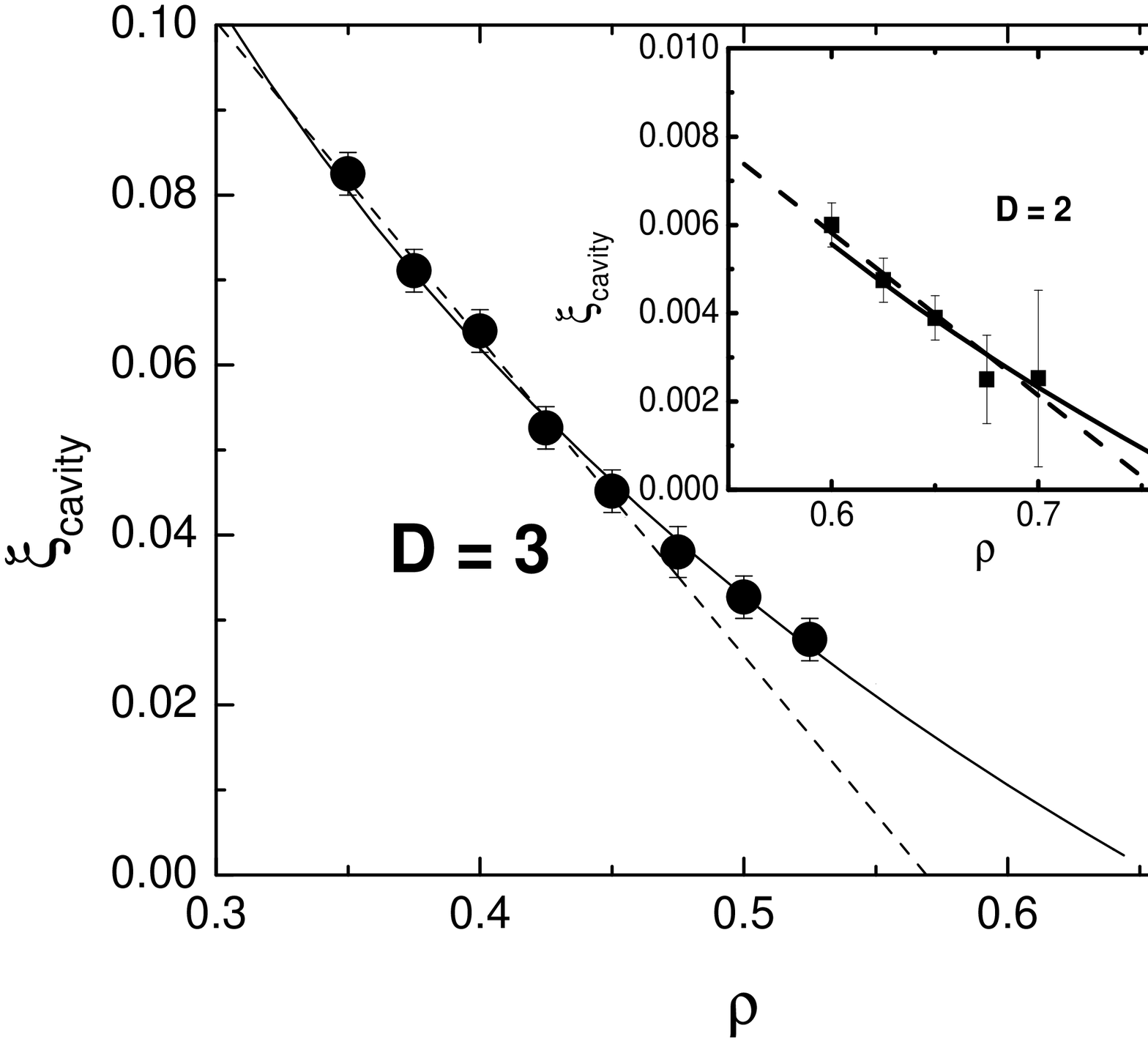, width=6in}}
\vspace{-25 mm}
\caption{}
\label{fig3b}
%\end{center}
\end{wrapfigure}

Fig. \ref{fig3} shows our results for $\xi_{\rm rattle}$ for a range of $\rho$ in both
$D$=3 and $D$=2.
We also plot the experimental cage sizes obtained from the dynamics of hard spheres 
to demonstrate the connection between
equilibrum properties and dynamics in these situations
\cite{cagingweeks,cagingheuer}. As a first step towards understanding the density dependence of $\xi_{\rm rattle}$, we
note that
in a simple ``cell" model of a dense homogeneous fluid, the 
cage size, $\xi$, is comparable to the average interparticle distance, so that
$\xi$ scales as $\rho^{- 1/D}$ at low densities and goes to 0 in the limit of
close packing.
%We have employed literature values of the random close packed density, i.e.,
%$\rho_{\rm rcp}$=0.644 ($D$=3) and 0.83 ($D$=2) \cite{speedyjam}.
In Fig. \ref{fig3} we compare $\xi_{\rm rattle}$ to the interpolation formula,
$\xi\propto\left [ \left(\frac{\rho_{rcp}}{\rho}\right)^{1/D}-1 \right]$
connecting these two limits. We employed this formula with
proportionality
constants
$\alpha$=0.87 ($D$=2), and $\alpha$=1.4 ($D$=3) and literature values of
$\rho_{\rm rcp}$= 0.83 ($D$=2) and 0.644 ($D$=3), respectively. These  results are 
consistent with the notion that this molecular measure of free volume vanishes in the limit
of random closed packing. Alternate forms often employed for studying
dynamic properties, such as $\xi_{\rm rattle}\sim (\rho_{\rm rcp}/\rho-1)^{\gamma}$
or $\xi_{\rm rattle}\sim(1-\rho/\rho_{\rm rcp})^{\gamma}$, also provide adequate fits to these
data, reiterating this fact. %The theoretical descriptions of $y(r)$ suggest that
%$\xi_{\rm rattle}= 2(1-\rho)^3/3\rho(2-\rho)$ in $D$=3 for $\rho\le$0.495, a result
%that is in reasonable agreement with simulation data in this range \cite{henderson}. 
%Since the equation of
%state of hard core fluids have been proposed to change at the crystallization density \cite{torquato,torquato2},
%the form of the cavity function, $g_{\rm disp}(r)$, and hence the rattle size, $\xi_{\rm rattle}$,
%might be expected to show different density dependences above and below the crystallization
%density. These facts provide some rationalization for extrapolation of this limited set of data alone.
The cavity size,
$\xi_{\rm cavity}$, also follows the same density dependence with $\alpha$=0.345 ($D$=3) [Fig. \ref{fig3b}].
However, for $D$=2
we had to fit both $\alpha$=0.038 and $\rho_{rcp}$=0.79 to obtain good agreement with
simulations, due to the large error
bars associated with this small quantity. Note that the number of data points for $\xi_{\rm cavity}$
are typically different from $\xi_{\rm rattle}$ due to sampling issues that make the
evaluation of $\xi_{\rm rattle}$ difficult at high densities.

\begin{wrapfigure}{c}{7in}
\vspace{30 mm}
\mbox{\epsfig{file=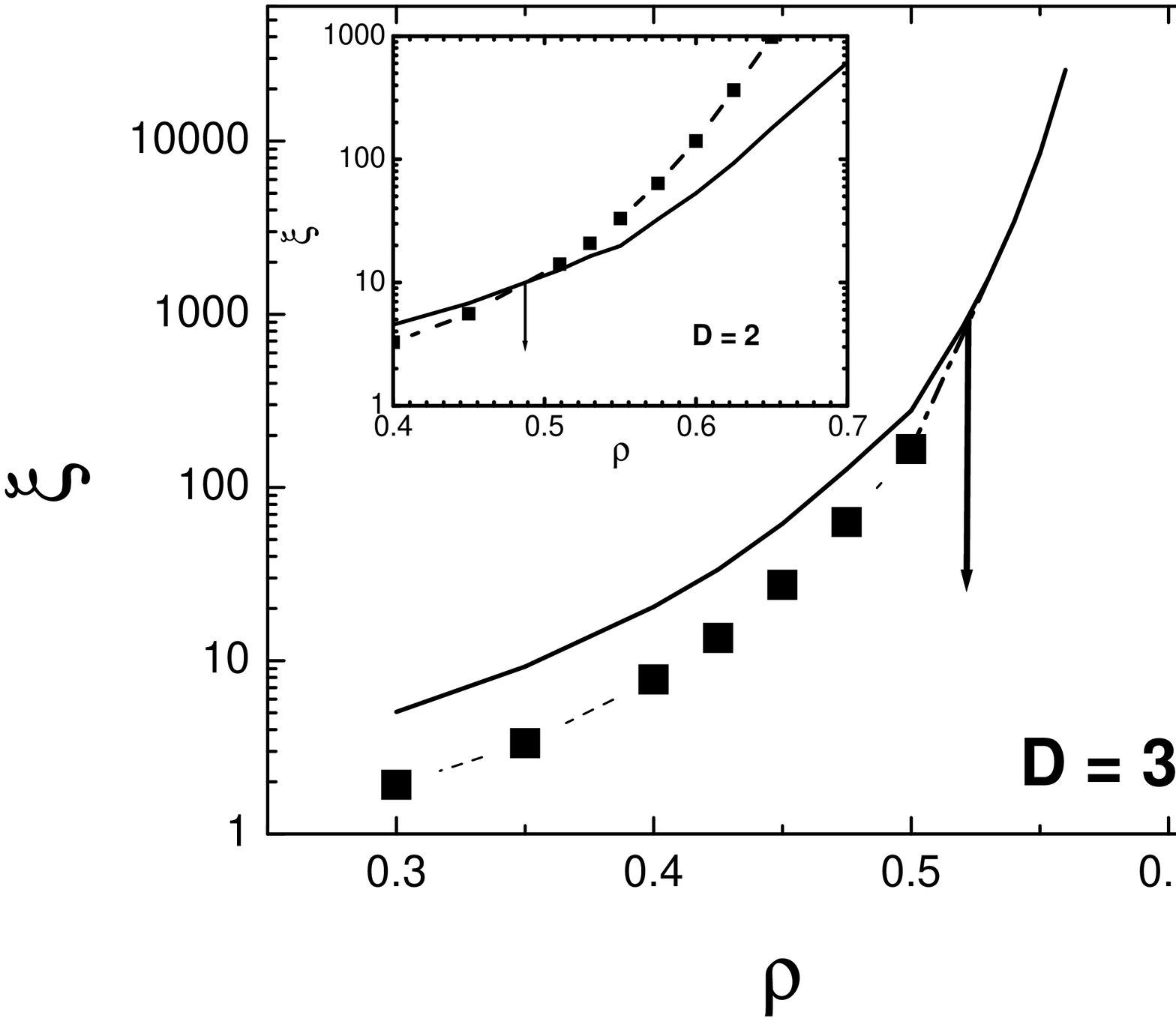, width=6in}}
\vspace{-40 mm}
\caption{}
\label{fig4}
%\end{center}
\end{wrapfigure}

These plots exhibit the usual problem that the value of 
$\rho$ where $\xi_{\rm rattle}$ and
$\xi_{\rm cavity}\rightarrow$ 0 depends sensitively on how we extrapolate to
higher densities. To illustrate this point, we note that
both $\xi_{\rm rattle}$ and $\xi_{\rm cavity}$ vary nearly
linearly with $\rho$ for 0.3$\le \rho<0.45$ for $D$=3, and
for 0.5$\le\rho<0.65$ for $D$=2.  Curiously, linear extrapolations of the low
density $\xi_{\rm rattle}$ and $\xi_{\rm cavity}$ data to zero
yield $\rho_G$=0.56$\pm$0.03 for $D$=3 and 
$\rho_G$=0.74$\pm$0.05 for $D$=2 \cite{speedyjam}. 
The density estimate in $D$=3 closely 
matches the reported experimental glass transition density, $\rho_G\approx 0.56-0.58$
presented by \cite{silescu,alder,woodcock}, where
diffusion coefficient data were extrapolated to zero over 
the restricted range $\rho\le$ 0.52.
The significance of this coincidence is unclear, since 
experiments and simulations where higher density data are 
considered (e.g., \cite{russel,woodcock,weitz,chaikin}) suggest that the viscosity does not
diverge at $\rho_G$, but rather at a density close to $\rho_{\rm rcp}$. 
Similarly, we find
that a plot of the logarithm of literature diffusivity data \cite{woodcock} varies linearly
with $\xi_{\rm rattle}^{-2}$, in agreement with the notion that random closed packing
is close to the point where the system ``vitrifies" \cite{starr}.
However, we note that the determination of the
exact density at which the viscosity diverges is somewhat uncertain due to the
inherent errors in the high density data and the extrapolation process. 
For two 
dimensional disks, the ``glass transition" density is harder to locate since most simulations
introduce polydispersity to prevent crystallization. 
For polydisperse disks there is the estimate $\rho_G\approx$ 0.8 \cite{doliwa,santhen}, but this
number is probably larger than for the monodisperse limit. Our $\rho_G$ estimate thus does
match the experimental estimates obtained from extrapolation of low density data.
Above we noted that visual examination of fluid configurations 
suggest that small, complex shaped cavities become 
increasingly separated with increasing $\rho$ (Fig. \ref{snapshot}).  
We next define a scale relevant to the typical separation of the ``void
clusters".  We expect that
spatial correlations associated with these void structures may have some relation to
dynamic heterogeneities in the dense fluid regime 
\cite{doliwa,sharon2,glotzer}, but the exact relationship is currently unclear.
Specifically, we consider the distance away from 
a given insertion point that we must go such that the probability 
of insertions in a spherical (circular in $D$=2) shell is larger than the peak 
probability of insertion, located in the immediate vicinity ($r<1$) of 
the first insertion point.  For example, Fig. \ref{fig1}b shows that this ``crossover" scale, 
$\xi_{\rm cross}$, at which insertions become more probable than 
in the immediate neighborhood occurs at $\approx 8$ for 
$\rho = 0.4$.  For randomly chosen insertions points, we expect the scale 
for successive insertions to be $\xi_{\rm rand} = (1/p_{\rm ins})^{1/D}$, which 
naturally increases for increased $\rho$.  [$p_{\rm ins}$ is the insertion probability
for a sphere into a snapshot of the fluid.] It is natural to then consider the 
role of correlations by comparing the behavior of 
$\xi_{\rm cross}$ and $\xi_{\rm ran}$.
Fig. \ref{fig4} shows both $\xi_{\rm cross}$ and $\xi_{\rm ran}$ as a function of $\rho$.  
At small $\rho$, $\xi_{\rm cross} < \xi_{\rm ran}$, apparently 
reflecting the tendency of particles to be clustered in the fluid (Fig. \ref{snapshot}). 
There is a characteristic $\rho$ where $\xi_{\rm cross}$ first exceeds $\xi_{\rm ran}$, 
and under these conditions we suggest that it is the holes that now cluster.
Notably, the density where $\xi_{\rm cross} = \xi_{\rm ran}$ corresponds 
to the density identified as the onset of ``caged" dynamics in both $D$=2 ($\rho\approx$0.48)
\cite{markus} and $D$=3 ($\rho\approx$0.53) \cite{cagingheuer,silescu}.  
Hence, we tentatively identify the point where the holes cluster as the onset of 
complex, structured fluid behavior, and possibly a structural indicator of the onset of dynamical heterogeniety.

It is useful to contrast our work to Ref. \cite{starr}, which has shown that the
rattle volume vanishes in the vicinity of the Vogel temperature for Lennard-Jones particles. 
For hard spheres, our results suggest that random close packing plays the role of the
``Vogel" point, i.e., the density where the viscosity diverges and the free volume extrapolates to
zero. Additionally, an alternate measure
of the range of structure provides an estimate of the onset of caging.

We acknowledge the National Science Foundation , Division of Materials Research for funding,
and Ralph Colby, Pablo Debenedetti and Robin Speedy for useful discussions.

\newpage
%\bibliography{glass}

\newpage
\begin{center}
{\bf Figure Captions}
\end{center}
\begin{enumerate}
{\item (a) Typical snapshots of the $D$=2 fluid. We show small portions of the simulation
cell for ease of visualization, and the circles drawn have a radius of 1. Thus, the
white spaces in these figures represent areas where centers of additional disks
can be placed without overlap. The left snapshot is for a density $\rho$=0.4, while
the right picture is at $\rho$=0.53.
(b) Main plot: $p_{\rm disp}(r)/4\pi$
as a function of $r$ for the three dimensional hard sphere fluid at $\rho$=0.4. The two arrows represent characteristic sizes: the one closest to the
origin is $\xi_{\rm rattle}$, while the other distance is $\xi_{\rm cross}$.
The $p_{\rm disp}(r)$ is used
to generate $y(r) \equiv
\frac{p_{\rm disp}(r) \Gamma(D/2)}{2\pi^{D/2} r^{D-1}}$, where $\Gamma(x)$ is
the gamma function, and $2\pi^{D/2}r^{D-1}/\Gamma(r) dr$ is the volume 
of a D-dimensional hypersphere shell of radius $r$ and thickness $dr$. The distance and
density dependence of $y(r)$ for $\rho\le$0.495 for $D$=3 are
consistent with past results \cite{henderson}.
Inset: Plot of $p_{\rm disp}(r)/4\pi$
as a function of $r$ for the three dimensional hard sphere fluid at a variety
of densities as sketched in the figure.}
\item{Plots of $\xi_{\rm rattle}$ as a function of $\rho$ for $D$=2 (squares) and
$D$=3 (circles). Uncertainties correspond to standard deviations in this
data as obtained from block averages. 
The triangles are experimental cage sizes for hard spheres as
reported by \cite{cagingweeks}. Here we have normalized
the experimental cage sizes by the bead diameter. Since our definition of
$\xi_{\rm rattle}$ may be expected to be smaller than the cage size by a factor of 2,
we divide the experimental data by 
2 to ensure a proper comparison. The full line is the best fit to the cell
model form $\xi_{\rm rattle}=\alpha \left [\left (\frac{\rho_{\rm rcp}}{\rho}
\right)^{1/D}-1 \right]$,
while the dotted line is a linear fit to lower density data as discussed in the text.}
\item{Cavity size, $\xi_{\rm cavity}$ plotted as a function of density. Main plot
is for hard spheres and the inset is for disks. Uncertainties are standard deviations
in the data. The dotted lines are best fits over the low density ranges
as discussed in the text, while the 
full line is the best fit to the cell
model form $\xi_{\rm cavity}=\alpha \left [\left (\frac{\rho_{\rm rcp}}{\rho}
\right)^{1/D}-1 \right]$.}
\item{Data for the $\xi_{\rm cross}$ (squares) and $\xi_{\rm ran}$ (lines) for hard spheres
[main plot] and disks [inset] as a function of density.}
\end{enumerate}
\newpage

\newpage

\newpage
\newpage

\end{document}